\documentstyle[aps,prl,multicol,epsf]{revtex}

\newcommand{\bleq}{\ifpreprintsty
                   \else
                   \end{multicols}\vspace*{-3.5ex}{\tiny
                   \noindent\begin{tabular}[t]{c|}
                   \parbox{0.493\hsize}{~} \\ \hline \end{tabular}}
                   \fi}
\newcommand{\eleq}{\ifpreprintsty
                   \else
                   {\tiny\hspace*{\fill}\begin{tabular}[t]{|c}\hline
                    \parbox{0.49\hsize}{~} \\
                    \end{tabular}}\vspace*{-2.5ex}\begin{multicols}{2}
                    \fi}
\newcommand{\bcols}{\ifpreprintsty\else\begin{multicols}{2}\fi}
\newcommand{\ecols}{\ifpreprintsty\else\end{multicols}\fi}

\begin{document}

\title{Charge Fluctuations in a Quantum Dot with a Dissipative 
Environment}
  
\draft

\author{Alex Kamenev$^a$, and Yuval Gefen$^b$ }

\address{
  $^{a}$Institute for Theoretical Physics University of California
Santa Barbara, CA  93106-4030, U.S.A.\\
  $^{b}$Department of Condensed Matter Physics
The Weizmann Institute of Science 76100 Rehovot, Israel\\
{}~{\rm (\today)}~ \medskip \\ \parbox{14cm} 
{\rm
We consider a multiple tunneling process into a quantum dot capacitively 
coupled to a dissipative environment. The problem is mapped onto an
anisotropic Kondo model in its Coulomb gas representation. The tunneling 
barrier resistance and the dissipative resistance of the environment 
correspond to the transverse and the longitudinal Kondo couplings respectively.
We thus identify a line in the parameter space of the problem which corresponds 
to a zero--temperature Berezinskii--Kosterlitz--Thouless like phase transition. 
The physics of coupling to the environment is elucidated and experimental 
consequences of the predicted transition are discussed.    
 }\bigskip \\ }

\maketitle

\bcols

\section{Introduction}
\label{s1}

The problem of the Coulomb blockade in zero dimensional quantum dots has
been introduced theoretically and studied both theoretically and
experimentally over the past decade \cite{BenJacob85,Mullen88}. 
First studies employed an
orthodox model within which the effect of charging has been accounted for
classically. More recent studies have included various quantum mechanical 
aspects
of the Coulomb interaction on various levels of rigor. Thus the
Altshuler--Aronov zero bias anomaly \cite{Altshuler79} 
has been related to the Coulomb 
blockade \cite{Panukov88,Nazarov89,Levitov96,Kamenev96}. In a slightly 
different language it was also realized 
that the coupling to a dynamically 
polarizable environment may modify the system's behavior qualitatively 
\cite{Devoret90,Girvin90}. The instantaneous tunneling of an electron into 
the dot leads (through the Coulomb interaction) to a shake--up excitation of 
the low--energy modes of the environment. This process, which is very similar 
in nature to the X--ray edge singularity (see Refs.  
\cite{Devoret90,Girvin90} and references therein), results in a power law 
current--voltage characteristics of a single tunneling barrier connected 
to a dissipative circuit.     

Another remarkable sequence of developments is associated with the study of 
the quantum fluctuation of the charge on the dot, or multiple tunneling 
processes  
\cite{Guinea86,Glazman90,Matveev91,Zwerger93,Grabert94,Shoeller94,Golubev94}. 
The importance of such processes  
in the vicinity of transmission resonance has been first realized 
by Glazman and Matveev \cite{Glazman90}. Later on Matveev \cite{Matveev91}  
mapped the problem onto the anisotropic Kondo model 
\cite{Yuval70,Anderson70}. This 
mapping is based on a projection of the Hamiltonian onto two charge states of
the dot. The two degenerate charge states of the dot (which corresponds, say, 
to $N$ and $N+1$ electrons respectively), are mapped onto two spin states, 
``up'' and ``down'', with the
coupling to the leads playing the role of a coupling to a Fermi sea. 
A small deviation from resonance conditions plays the role of
a constant magnetic field in the Kondo problem, leading to
Zeeman splitting. Extensions of this problem to other variants have been
carried out employing equivalent methods \cite{Grabert92}.

\par 
In the present paper we consider the effect of a dissipative environment on  
the quantum fluctuations of charge of the dot. 
This relatively straightforward 
combination of previously studied effects leads to a qualitatively novel 
physical situation, which, to the best of our knowledge, has not been 
discussed before  in the present context. 
The dissipative resistance of the environment, $r$, 
capacitively coupled to the dot, 
gives rise to a new parameter, important for the physical description
of the system. In the language of the 
anisotropic Kondo model \cite{Yuval70,Anderson70} 
this new parameter is precisely equivalent to a (ferromagnetic) 
$z$--axis coupling, $J_z$. Let us recall that the resistance of the tunneling 
barrier, $r_t$, is mapped   \cite{Matveev91} onto the (inverse) $x$-$y$ 
coupling,
$J_{\perp}$. As a result one obtains a separatrix line in the parameter space 
($r_t$-$r$) of the system. Crossing this line by varying
the values of the respective resistances  
leads 
to a zero--temperature phase transition. In the ``ferromagnetic'' phase the 
system is overdamped and its behavior is akin to that of a single tunneling 
barrier connected to a dissipative circuit \cite{Devoret90,Girvin90}. Namely,
the peak conductance decreases as a certain power of temperature. By contrast,
in the ``anti-ferromagnetic'' phase, the peak hight undergoes re-entrance 
behavior   and eventually approaches a perfect conductance limit 
\cite{Matveev91}
(for a single channel tunneling contact).  

Technically we proceed by mapping  
the problem of multiple tunneling into the dot, onto that of a
classical neutral gas in one dimension with long range
interactions. In the past, such a mapping has been employed in the context
of the anisotropic Kondo problem 
\cite{Yuval70,Anderson70} and  dissipative quantum tunneling 
\cite{Chakravarty82}.
In the latter context the analogous phase transition
corresponding to localization--delocalization in double quantum well was 
exploited extensively. The quantum dot realization of this transition, 
proposed here, may be the most suitable for experimental realization. 
A related, though qualitatively different
 gas, which also can be realized in the present context, 
has been employed in the study of a one--dimensional
Luttinger model with an impurity \cite{Kane92}. 
Here we restrict ourselves to a 
study of a quantum dot connected to a lead through a 
single tunnel barrier, occasionally referred to 
in the literature as a ``single electron 
box''. We discuss the mapping onto a classical gas and the subsequent 
renormalization group (RG) procedure on the level of 
the partition function of such 
a ``box''. A quantity more directly related to the experiment is the 
conductance through a dot with the two barriers \cite{Shoeller94,Golubev94}. 
We believe, though,  
that our calculations reflect the experimental situation with $r_t$ being the 
resistance of the weakest of the two barriers.  
We also require that our dot is large enough, in a sense that one 
may disregard the discreteness of its single particle spectrum.

\par 
The outline of this paper is the following. In Section \ref{s2} we define
our problem in terms of a model Hamiltonian and an equivalent electrical
circuit, and discuss qualitatively the effect of the environment. 
We also discuss a complete solution of the problem in  
lowest order in the tunneling amplitude.   
Mapping the quantum fluctuations of the charge onto a classical Coulomb gas is
presented in Section \ref{s3}. The fully and partially screened cases are 
solved using a renormalization group procedure.  
In Section \ref{s4} we discuss the physics of our results and briefly comment 
on possible generalizations. Technical details of the derivation are 
presented in two Appendices.

\section{Model Hamiltonian and an Equivalent Circuit}
\label{s2}

We consider a small capacitance dot, described by the Hamiltonian
\begin{equation} 
{\cal H}_0=\sum_\alpha\epsilon_\alpha a^+_\alpha a_\alpha +{V^{[0]}\over
2}\left( \sum_\alpha a^+_\alpha a_\alpha-{\cal N}_0\right)^2\, ,
                                                              \label{e1}
\end{equation}
where $a^+_\alpha (a_\alpha)$ is a creation (annihilation) operator of an
electron in an exact single particle state $\alpha$ (the latter is defined
including disorder potential and spin); 
$V^{[0]}/2=e^2/2C$ is the charging energy,
associated with the total charge on the dot ($C$ is the dot's
self--capacitance);  ${\cal N}_0$ is the effective charge of the positive
background.  This may be regarded as the large wave--length ($q$=0)
component of the interaction. Other, $q\ne 0$, components are ignored.
The dot is assumed to be weakly 
coupled to a conducting lead, described by a gas of non--interacting
electrons
\begin{equation} 
{\cal H}_{lead}=\sum_k\epsilon_k d^+_k d_k\; ,
                                                            \label{e2}
\end{equation}
where $d_k,d^+_k$ are Fermi operators. The coupling to the lead is given
by
\begin{equation} 
{\cal H}_{coupling}=\sum_{k,\alpha}\big(W_{ka} d^+_ka_\alpha
+W^*_{k\alpha}
a^+_\alpha d_k\big)\; .  
                                                             \label{e3}
\end{equation}
Below we shall consider a single point contact and impose
$W_{k\alpha}=W$ \cite{foot0}. The parameter $W$ is related to the bare (high
temperature) tunneling resistance, $R_t= (2\pi\hbar/e^2) r_t$, where 
\begin{equation} 
(r_t)^{-1}=(2\pi)^2 |W|^2\nu^{[0]}\nu^{[l]}\; , 
                                                             \label{e4}
\end{equation}
where $\nu^{[0]}$ and $\nu^{[l]}$ are the (bare) density of states (DOS) of
the dot and the lead respectively. Finally, we
shall assume that the dot is embedded in an electromagnetic environment,
felt by the dot as an external noise. This noisy background is described as
an effective time dependent term in the Hamiltonian 
\begin{equation} 
{\cal H}_{noise}=\eta(t)\sum_\alpha a^+_\alpha a_\alpha\; . \label{e5}
\end{equation}
Below we specify the noise, $\eta(t)$, in terms of the environment's
characteristics. Equivalently, we could say that the dot is connected to a 
bath of harmonic oscillators which
are integrated out. The  spectral density of $\eta(t)$ reflects the nature of
the environment \cite{Chakravarty82}. In this sense the averaging over noise 
should be understood as annealed and not as quenched. 
The total effective Hamiltonian of the system is thus
\begin{equation}  
{\cal H}= {\cal H}_{dot} +{\cal H}_{lead}+ {\cal H}_{coupling} +{\cal
H}_{noise}\; . 
                                                           \label{e6}
\end{equation}

\par 
The physical effect of the environment \cite{Devoret90,Girvin90} 
can be understood in
the following manner. Imagine that an electron has been injected into the
dot. The total effective charge, 
$e(\langle N\rangle-{\cal N}_0)$, interacts with
charges in the environment, leading to their redistribution (polarization).
The polarized charge of the environment reduces the energy cost of adding 
(or removing) an electron to (or from) the dot. There is a certain 
finite time scale characterizing this redistribution of the 
environmental charge, rendering the effective interaction in the dot
non--instantaneous (retarded) \cite{Spivak,Levitov96}. 
This time constant can be described as
an RC time of an effective electric circuit. It might appear that the
results obtained through the ensuing analysis are highly non-universal and
depend on the particular choice of the model for the environment. We
stress that the effect of the environment is quite general, 
the only model dependent feature being the
concrete form of the screened zero--mode interaction, $V(\omega)$ (see below).
It enters through the dependence of $V(\omega)$ on an effective 
impedance of the environment \cite{Schon89}, and
is due to the fact that details of the environment's polarization may
depend on various objects located far from the dot.

\par
\begin{figure}
\vglue 0cm
\hspace{0.01\hsize}
\epsfxsize=0.9\hsize
\epsffile{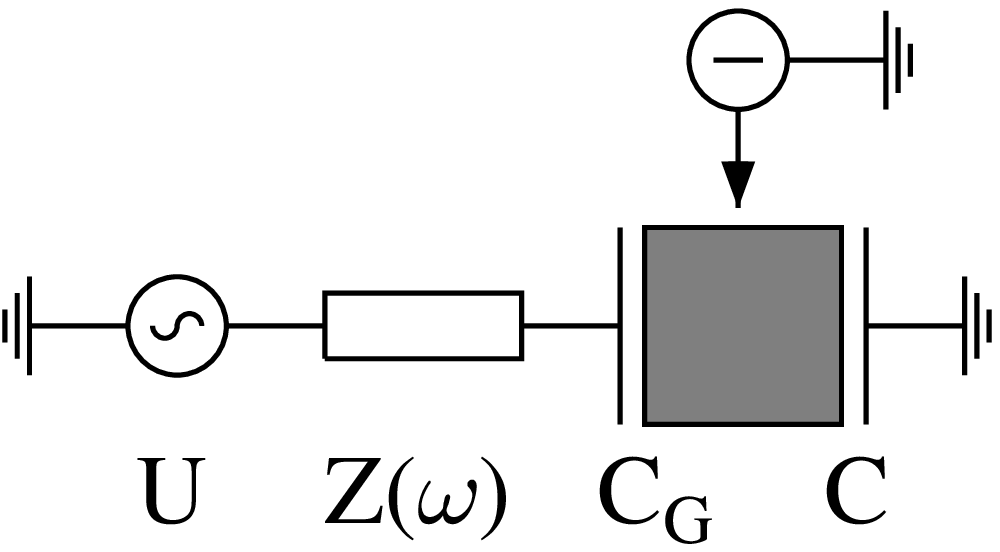}
\refstepcounter{figure} \label{fig1}
{\small FIG.\ \ref{fig1}. An equivalent circuit describing an experimental
setup of an electron
injection into a dot. For details see the text. \par}
\end{figure}

 
Fig. \ref{fig1} shows a simplified equivalent circuit depicting an experiment
of an electron injection into a quantum dot. The dot is assumed to be weakly 
coupled to the lead; $C$ represents the self--capacitance of the dot. 
The dot is
capacitively coupled (through $C_G$) to a dissipative environment whose
impedance is $Z(\omega)$. The source $U$ represents the 
equilibrium voltage noise
of the environment.

\par 
In the lowest order in the tunneling amplitude, $|W|^2$, the  conductance of the 
tunneling junction is fully determined by the DOS of the 
dot. The latter may be obtained through the imaginary--time one particle Green
function of the Hamiltonian ${\cal H}_0+{\cal H}_{noise}$. 
It is, of course,  strongly modified by the effects of interaction and 
coupling to the environment. For completeness we briefly reproduce the
calculations of Ref. \cite{Kamenev96} in Appendix A, which lead  
to the following expression 
for the one particle Green function 
\begin{equation} 
{\cal G}_\alpha (\tau) ={\cal G}^{[0]}_\alpha (\tau)
e^{-S(\tau)}\; . 
                                                          \label{e7}
\end{equation}
Here the effective action, $S$, is given by 
\begin{equation}
S(\tau)=T\sum_{m\neq 0}\frac{V(\omega_m)}{\omega_m^2}
(1-e^{i\omega_m\tau})\, ,
                                                          \label{e9}
\end{equation} 
where $T$ is the temperature and $\omega_m$ are the bosonic Matsubara
frequencies, $\omega_m\equiv 2\pi mT$. The form of the screened
interaction, $V(\omega_m)$, can be easily deduced relying on the 
equivalent circuit picture, Fig. \ref{fig1}. It is shown in Appendix A  that
\begin{equation}
V(\omega_m)=V^{(0)}- K(\omega_m)\, ,
                                                          \label{e10}
\end{equation}    
where $K$ is the correlator of the Gaussian noise felt by a dot. 
In the imaginary time domain
$K(\tau-\tau')=\langle\eta (\tau)\eta (\tau')\rangle$,  
and may be evaluated
employing the fluctuation--dissipation theorem. Consulting Fig. \ref{fig1} the
correlator of the total noise in the equivalent circuit (in terms  of real 
frequencies) is: 
$ \langle UU\rangle_\omega =e^2 i\omega Z_{tot}(\omega)\,$, 
where the total impedance of the equivalent circuit is
$Z_{tot}=(i\omega C)^{-1} +(i\omega C_G)^{-1}+Z(\omega)\;$ .             
The corresponding voltage drop on the dot is 
$\eta = U(i\omega C)^{-1}/Z_{tot}\, $.
Substituting the above given expressions in Eq.~(\ref{e10}) and rewriting it 
in a finite temperature form one obtains
\begin{equation}
V(\omega_m)=\frac{e^2}{C}\, \, 
\frac{C/C_G+|\omega_m| ZC}{1+C/C_G+|\omega_m| ZC}\, . 
                                                          \label{e16}
\end{equation} 
In the high frequency limit the interaction is unscreened, whereas in the 
low frequency limit we obtain a (partially) screened
interaction, where $C$ is replaced by $C+C_G$. The scale of the 
crossover frequency 
is given by $\Omega^{-1}=|Z(\Omega)|C$ \cite{foot1}.

Coming back to the action, Eq. (\ref{e9}), the long time behavior of $S(\tau)$
is given by the small frequency limit of $V(\omega_m)$, yielding
\begin{equation} 
S(\tau)\approx |\tau|{e^2\over 2(C+C_G)}\,\,\,\, \mbox{for}\,\,\,\, 
ZC_G\ll\tau\ll\beta\;.
                                                      \label{e19}
\end{equation}
This is the case of a partially screened interaction, 
$C\rightarrow (C+C_G)$. One
expects that in this case the addition of charge to the dot costs finite
energy even at long times, which is why Eq. (\ref{e19}) is linear in $\tau$. This
immediately implies the existence of a gap in the DOS, which is a manifestation of the Coulomb 
blockade. In the special case of a fully screened interaction
$(C_G\rightarrow\infty)$, the long time interaction vanishes, i.e. we expect
$V(\omega_m\rightarrow 0)=0$. This is the case of a single tunneling 
contact in series with a 
classical impedance, $Z$, considered in Refs. \cite{Devoret90,Girvin90}. For a
purely Ohmic environment, $Z=R$, the interaction is
\begin{equation} 
V(\omega_m)=V^{[0]} {|\omega_m|\over\Omega+|\omega_m|} \; ,
                                                           \label{e21}
\end{equation}
with $\Omega\equiv (RC)^{-1}\;$.
On short time scales ($\omega>\Omega$) before the environment becomes fully
polarized to screen out the extra charge injected into the dot, the
charging energy is finite. Thus the injection of an electron can
be considered as ``tunneling under the barrier"  in the time
direction. The very same physics has been discussed in the context of
tunneling into two--dimensional systems with diffusive disorder 
\cite{Spivak,Levitov96}. 
The related energy cost on short time scales suppresses free particle exchange
between the dot and the particle reservoir (although Coulomb blockade  in
its strict sense is absent). This leads to a suppression of the DOS, 
hence to a zero bias anomaly. Substituting Eq. (\ref{e21}) 
in Eq. (\ref{e9}) one
obtains for $\Omega^{-1}\ll\tau\ll\beta$
\begin{equation} 
S(\tau)\approx 2r\,\ln (\Omega |\tau|) \;,
                                                      \label{e23}
\end{equation}
where 
\begin{equation} 
r\equiv {V^{(0)}\over 2\pi\Omega} = {R\over (2\pi\hbar/e^2) }
                                                     \label{e24} 
\end{equation}
is the dimensionless resistance of the environment. 
This form of $S$ leads to a power
law DOS \cite{Kamenev96} and hence power law current--voltage characteristics  
\cite{Devoret90,Girvin90}.

\begin{figure}
\vglue 0cm
\hspace{0.01\hsize}
\epsfxsize=0.95\hsize
\epsffile{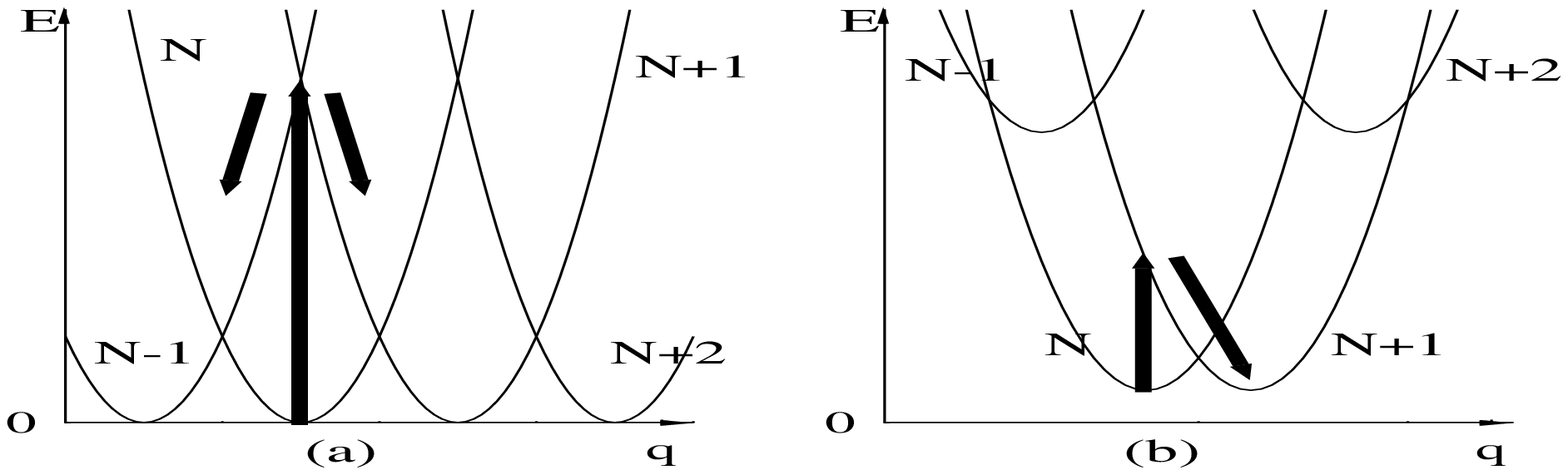}
\refstepcounter{figure} \label{fig2}
{\small FIG.\ \ref{fig2}.   
Energy of the dot with $N$ particles as function of the gate capacitance 
charge, $q$. Instantaneous tunneling of an electron into the dot, 
and subsequent slow
relaxation of the charge are indicated by arrows. (a)~~Full screening,
$C_G=\infty$.
(b)~~Partial screening, a resonance scenario.}
\end{figure}%

\par 
Before concluding this section we draw a qualitative picture of the
difference between the fully and partially screened
interaction scenarios. Consider first the case of $C_G\rightarrow\infty$ and
denote
the number of electrons on the dot by $N$. At long times the environment
polarizes,
such that the total neutralizing positive charge (including both the original
positive 
background on the dot, ${\cal N}_0$, and the accumulated charge on the gate
capacitance, $q$) exactly balance off the charge $N$. Fig. \ref{fig2}a  depicts
the charging energy of the dot as function of $q$, for various values of
$N$. This is given by $E_N(q)= e^2(N-{\cal N}_0-q)^2/2C$. The
fact that the center of each parabola corresponds to $E_N=0$ is a
manifestation of perfect screening. Upon adding an electron to the dot
$(N\rightarrow N+1)$, the charging energy is initially increased, but at
later times $q$ relaxes towards the optimal value corresponding to $N+1$.
A very different situation arises for a partially screened interaction,
$C_G<\infty$. The charging energy is now given by 
$E_N(q)= e^2(N-{\cal N}_0-q)^2/2C+e^2 q^2/2C_G$.
The curvature of each parabola 
corresponds to $e^2/2C+e^2/2C_G$, whereas the centers of the various 
parabola are located on 
the parabola $e^2/2(C+C_G)$. Adding an electron to
the system initially costs some energy (as in
the previous case), but eventually the system relaxes to the partially
screened state whose energy, although less than in the first moment 
after tunneling, is not in 
general equal to that of before tunneling event.  
By tuning ${\cal N}_0$  it is possible to have at most two states having
the same final energy. Exactly this resonance scenario (with
${\cal N}_0=N+{1\over 2}$) is depicted in Fig. \ref{fig2}b. 
This picture will be useful while considering multiple tunneling events, 
or quantum  fluctuations of charge on the dot.

\section{Quantum fluctuations of charge}
\label{s3}

Our aim now is to address the full Hamiltonian, Eq. (\ref{e6}), including the possibility of 
multiple tunneling events. To simplify calculations we first consider a partition function 
of the system. We formally exactly  expand it in powers of ${\cal H}_{coupling}$ and 
calculate separately each term of the expansion. Details of this procedure may be found in 
Appendix B. As a result one obtains the normalized partition function of the dot coupled 
to the lead in the following form 
\bleq 
\begin{equation}
{Z\over Z^{[0]}Z^{[l]}}
=\sum^\infty_{M=0}{\theta^{2M}\over M!M!} 
\int\limits^\beta_0\!\!\ldots \int\limits^\beta_0\!\!     
d(\Omega\tau_1) \ldots d(\Omega\tau_{2M}) 
\exp \left\{ \sum\limits^{2M}_{i<j} 
(-1)^{i+j} \left[ 2\ln(\Omega|\tau_i-\tau_j|)+S(\tau_i-\tau_j)\right]\right\} ,
                                                                 \label{e50}
\end{equation}
\eleq
where $Z^{[0]}$ and $Z^{[l]}$ are partition functions of uncoupled dot and lead 
correspondingly, $S(\tau)$ is defined by Eq. (\ref{e9}) and 
\begin{equation} 
\theta \equiv
\sqrt{|W|^2\nu^{[0]}\nu^{[\ell]}} = {1\over 2\pi}\sqrt{1\over r_t}\;\; .
                                                                \label{e51} 
\end{equation} 
The r.h.s. of Eq. (\ref{e50}) 
may be considered as the grand-canonical partition function of a
{\it classical} one--dimensional {\it neutral} gas consisting of positive and
negative charges which interact through the potential $2\ln \Omega |\tau|+S(\tau)$. The
quantity $\theta$, related to the resistance of the tunneling barrier by 
Eq. (\ref{e51}) 
(cf. Eq. (\ref{e4})), represents the {\em fugacity} of the gas.  
The interaction potential of the neutral gas deserves some elaboration. 
The first term, $2\ln \Omega |\tau|$, came from the Green functions of 
non--interacting 
electrons in the dot and the lead. At finite temperature one should substitute 
$\tau\rightarrow \sin(\pi\tau/\beta)\beta/\pi$. One should also understand 
that the 
interaction potential given above is for large separation between charges 
($\Omega\tau\gg 1$), for $\tau\rightarrow 0$ it should vanish. 
Note that $S(\tau)$ obeys this property as well as periodicity with $\beta$ 
(by definition, cf.  Eq. (\ref{e9})).  

Before treating specific examples we make a few comments. Our gas which consists of
the same 
number of positive and negative charges is fundamentally different from the
logarithmically interacting gas known from the theory of random matrices \cite{Mehta91}.
The
latter contains charges of the same sign placed in a confining potential. In
our case
the positions of the charges along the one dimensional (imaginary) time axis 
correspond to the times of the (instantaneous) tunneling events: positive
charges --
tunneling into the dot; negative -- tunneling out of the dot. In general, any
ordering
of the charges, compatible with global neutrality is permissible. Neutrality
follows from the fact that the partition function is given by a trace, thus
forcing the initial and final charge states to coincide. At finite
temperatures the interaction becomes periodic over the period 
$0\leq\tau<\beta$. 
The resulting gas 
is thus defined on a ring of circumference $\beta$.

\subsection{Fully Screened Interaction}
\label{s3a}

As has been discussed in Section \ref{s2}, the fully screened interaction scenario
corresponds to an infinite gate capacitance, $C_G\rightarrow\infty$, in the
equivalent circuit (cf. Fig. \ref{fig1}). This situation is equivalent to the case of a
circuit with a single tunneling barrier, considered in Refs. \cite{Devoret90,Girvin90} in 
lowest order in tunneling. We provide here a systematic treatment of higher 
order processes. 
Below we shall consider a purely Ohmic environment.  
The effective action to be substituted in the expression for the partition
function is given (for $(\Omega|\tau_i-\tau_j|\gg1)$ by Eq. (\ref{e23}). 
Then the partition function is 
\bleq
\begin{equation}
{Z\over Z^{[0]}Z^{[l]}}
=\sum^\infty_{M=0}{\theta^{2M}\over M!M!} 
\int\limits^\beta_0\!\!\ldots \int\limits^\beta_0\!\!     
d(\Omega\tau_1) \ldots d(\Omega\tau_{2M}) 
\exp \left\{ (2+2r)\sum\limits^{2M}_{i<j} 
(-1)^{i+j} \ln(\Omega|\tau_i-\tau_j|) \right\}\; .
                                                                 \label{e55}
\end{equation}
\eleq
This form breaks down when inter--charge distances are small, i.e. when the
density of the Coulomb gas is high. This is a partition function of a
classical 1d
gas of unit charges interacting through a 2d Coulomb (logarithmic) potential
at temperature $(2+2r)^{-1}$ and fugacity $\theta$. We denote this gas as {\it
non-alternating}, meaning that {\it any} neutral sequence of charges is
allowed. The correspondence between the quantum dot problem and the classical Coulomb gas 
is summarized in a Table 1. 
Mapping onto the same problem has been found in the context of dissipative
tunneling
in a periodic potential \cite{Chakravarty82} and more recently in the study of the Luttinger
liquid
with an impurity \cite{Kane92}. The analogy stems from the observation that the Coulomb
gas,
Eq. (\ref{e55}), may be obtained from the partition function of a 
one--dimensional
quantum bosonic model with an action
\begin{eqnarray} 
S[\varphi (x,\tau)]=&&{1\over 2+2r} \int\!
dxd\tau \varphi(\partial^2_\tau+\partial^2_x) \varphi
\nonumber \\
&&-2\theta\Omega\int\!
d\tau\,\cos\big(
\sqrt{4\pi}\varphi(x=0,\tau)\big)\;. 
                                                        \label{e56}
\end{eqnarray} 
After integrating the bosonic field at all points except $x=0$, one ends up
with the
effective action of the form
\begin{equation} 
S[\varphi(\tau)]=\int\!d\tau\left[{1\over
1+r}\varphi|\partial_\tau|\varphi -
2\theta\Omega\, \cos\left(\sqrt{4\pi}\varphi\right)\right]\;. 
                                                       \label{e57} 
\end{equation}
Exponentiating the action and expanding  in powers of $\theta$ one 
directly arrives at Eq. (\ref{e55}) where $\Omega$ plays the role of 
the high energy cutoff.

\begin{figure}
\vglue 0cm
\hspace{0.01\hsize}
\epsfxsize=1.1\hsize
\epsffile{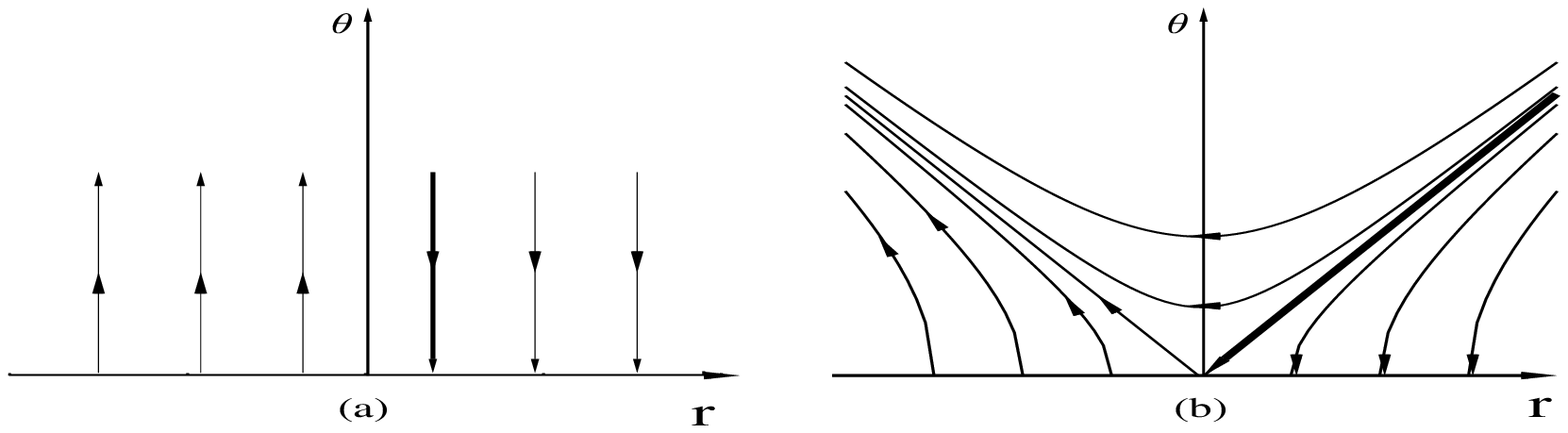}
\refstepcounter{figure} \label{fig3}
{\small FIG.\ \ref{fig3}. 
RG flows of the Coulomb gas: (a)~~ a non-alternating gas;
(b)~~an alternating gas. \par}
\end{figure}

We follow here an approach based on a renormalization group (RG) study of the
model \cite{Bulgadaev82,Kane92}. One
integrates out the high energy (short time) degrees of freedom by continuously
rescaling $\Omega\rightarrow\Omega/\xi$, where $\xi$ varies from 1 up
to
$\Omega/T$ ($\beta=T^{-1}$ is the maximal possible size of
the classical gas). 
As a result one arrives at the following well--known RG equations
\cite{Bulgadaev82,Kane92} 
\begin{mathletters}
\begin{eqnarray} 
&&{d\theta\over d\ln\xi}=-r\theta\;,
                                       \label{e58} \\
&&{dr\over d\ln\xi}=0\;.
                                         \label{e59}
\end{eqnarray}
\end{mathletters}
The resulting RG trajectories on the ($r$, $\theta$) plane are depicted 
on Fig. \ref{fig3}a. 
In the context of a Luttinger liquid with an impurity, $r$ represents the
strength
of the interaction and is therefore a feature of the bulk system. The fugacity
$\theta$ represents the strength of the point--like impurity. It is then clear
that,
while the interaction in the liquid may renormalize the impurity strength, the
local
impurity cannot renormalize the interaction strength in the bulk \cite{Finkelstein}. 
The separatrix 
$r=0$ corresponds to the presence of a Berezinskii--Kosterlitz--Thouless 
(BKT)  like phase
transition in a system of logarithmically interacting particles 
\cite{Kosterlitz74}. 
The line $r=0$
corresponds to the non--interacting system (cf. Eq. (\ref{e50}) with
$S=0$); evidently the tunneling transparency is not renormalized on this line.

\par 
For positive $r$ (which is the only physically meaningful scenario in the
present context) the fugacity $\theta\sim |W|$ is renormalized down, 
$\theta\sim\xi^{-r}$. 
This provides a justification for a lowest order (in $W$) perturbation theory at low
temperatures, which was employed in Section \ref{s2}. Indeed,
integrating Eq. (\ref{e58}) and assuming that the relation between fugacity and
resistance,
Eq. (\ref{e51}), is conserved in the process of renormalization, one obtains
\begin{equation} 
r_t\sim T^{-2r}\;.
                                                     \label{e60}
\end{equation}
This zero--bias anomaly behavior
has been derived previously  \cite{Devoret90,Girvin90,Kamenev96} in a 
calculation which was 
non--perturbative in interaction, but perturbative in the tunneling.  
A similar power-law
divergence of the resistance takes place in a repulsive Luttinger liquid with
a single impurity \cite{Kane92}.

\subsection{Partially Screened Interaction}
\label{s3b}

We now turn to a somewhat less trivial situation, that of a partially screened
interaction. This corresponds to a finite $C_G$ in Fig. \ref{fig1}.  
The ground state
energies corresponding to different numbers of electrons in the dot are
generically
not degenerate.  We refer though to a particular scenario where the gate
voltage
(effectively ${\cal N}_0$ in Eq.(\ref{e1})) is tuned such that the ground state
energies
corresponding to the charging states $N$, $N+1$ are degenerate, cf.
Fig. \ref{fig2}b.  
Injecting an electron to form any other charge state costs at least energy
equal to
$\tilde\Omega=e^2/2(C+C_G)$ at long time scale.  
If we set the temperature to be less than this value
one may neglect such states.  Consequently the charge on the dot can fluctuate
only
between $N$ and $N+1$. The corresponding classical model is that of a gas
of 
strictly alternating charges: a positive charge (tunneling into the dot) is
followed by a negative charge (tunneling out of the dot).  Similarly to the
fully
screened scenario discussed above, and following the same lines of derivation,
the
fact that the (two) available states are degenerate renders the long time
($\tau\gg\tilde\Omega^{-1}$)  interaction logarithmic.  
The coefficient in front of the logarithm is determined by the slope of
$V(\omega)$ at small $\omega$, which is given according to Eq.~(\ref{e16}) by  
\begin{equation}
\tilde r \equiv r\left[\frac{C_G}{C+C_G}\right]^2\;.
                                                     \label{e62}
\end{equation}
As a result, the normalized partition function takes the form (see Appendix B)
\bleq
\begin{equation}
{Z\over Z^{[0]}Z^{[l]}}
=\sum^\infty_{M=0}\theta^{2M}
\int\limits^\beta_0\!d(\tilde \Omega\tau_1)
\int\limits^{\tau_1}_0\!d(\tilde \Omega\tau_2) \ldots
\int\limits^{\tau_{2M-1}} _0\!d(\tilde\Omega\tau_{2M})
\exp\left\{ 
(2+2\tilde r) \sum\limits^{2M}_{i<j} 
(-1)^{i+j}\ln(\tilde\Omega|\tau_i-\tau_j|) \right\}\;.
                                                      \label{e63}
\end{equation}
\eleq

\par 
The very same Coulomb gas was first discovered by Yuval and Anderson in
the
context of the Kondo problem \cite{Yuval70}. In that case alternating positive and
negative
charges correspond to the up and down spin flips. The mapping between an
isolated
dot near the Coulomb resonance and the anisotropic 
Kondo problem was first presented in Ref. \cite{Matveev91}. 
There the coupling to the environment was absent ($\tilde r=0$) and the inverse 
temperature of the gas was exactly 2 (see Table 1), 
rendering the corresponding Kondo problem to be 
always anti-ferromagnetic. Here we have introduced a coupling to the dissipative 
environment, which may change the situation qualitatively by driving the system 
through anti-ferromagnetic--ferromagnetic transition. Note that a small
deviation from exact resonance conditions plays the role of Zeeman splitting
due to
a constant magnetic field in the Kondo model. In the case of the Kondo problem
both
the fugacity and the inverse temperature are properties of the local spin,
hence
both may be renormalized. The RG equations assume the form 
\cite{Anderson70,Kosterlitz74}
\begin{mathletters}
\begin{eqnarray}
&&\frac{d\theta}{d\ln\xi}=-\tilde r\theta\; ;
                                                     \label{e64}\\
&&\frac{d\tilde r}{d\ln\xi}=-4\theta^2\;.
                                                     \label{e65} 
\end{eqnarray}
\end{mathletters}
In the limit of extremely small fugacity, $\theta\ll|\tilde r|$, the gas is so
dilute
that the requirement of alternation does not play any role. Indeed in this
case the
RG equations of alternating gas coincide with Eqs. (\ref{e58}) and (\ref{e59}). 
To handle the case of finite fugacity one notices that
\begin{equation}
\tilde r^2-(2\theta)^2 = const\;.
                                                     \label{e66}   
\end{equation}
As a result the RG trajectories are hyperbola in the $(\tilde r,\theta)$
plane. The
RG diagram is depicted in Fig. \ref{fig3}b. 
The most striking feature of this diagram is the
presence of the separatrix at $\tilde r=2\theta$. In terms of the bare
parameters of
the problem the equation for the separatrix has the form
\begin{equation} 
\pi\sqrt{r_t}\, \tilde r=1\; .
                                                      \label{e67}    
\end{equation}
The low temperature behavior of the system is qualitatively different,
depending on
whether the bare parameters are chosen to be below or above the separatrix
line.

\par 
Let us first consider the system below the separatrix, $r_t<(\pi\tilde
r)^{-2}$. In this case the fugacity scales down, and the resulting Coulomb gas
is
dilute (the system is in the ``cold'' phase, i.e., overdamped by the environmental
resistance). Lowest order perturbation theory should be adequate while
renormalization provides some corrections. One can easily integrate 
Eqs. (\ref{e64}) and (\ref{e65}) by putting
\begin{equation}
\tilde r(\xi)=A\coth (A\ln\xi +B);\quad 
2\theta(\xi)=A/\sinh (A\ln\xi +B)\;,
                                                     \label{e68}    
\end{equation}
where $A^2\equiv\tilde r^2-(2\theta)^2$ and $\coth B\equiv\tilde r/A$. 
For $T\ll\tilde\Omega$ one obtains
\begin{equation} 
r_t(T)\sim T^{-2A}\;. 
                                                     \label{e69} 
\end{equation} 
If $\theta\ll\tilde r$, then $A\approx\tilde r$ and Eq. (\ref{e69}) coincides with
the
result of perturbation theory (in the tunneling strength), cf. 
also Eq. (\ref{e60}).
In the more general case the low temperature exponent is somewhat smaller than the
naive perturbation theory prediction, indeed $A < \tilde r$.

\par 
At the separatrix line, Eq. (\ref{e67}), the system undergoes a zero--temperature
BKT phase transition. Above the separatrix line,
$r_t>(\pi\tilde r)^{-2}$, 
the fugacity exhibits reentrance behavior and eventually is renormalized
up. The corresponding Coulomb gas is in the ``hot'' or plasma phase. Note that the
dot without a dissipative environment $(\tilde r=0)$ is always in the hot phase. 
This latter case has been
extensively studied since the work of Matveev \cite{Matveev91} 
(see e.g. Ref. \cite{Zwerger93}). The
temperature below which the reentrance behavior takes place is analogous to
the
Kondo temperature, $T_K$.  To find $T_K$ we integrate
Eqs. (\ref{e64}) and (\ref{e65}) by the substitution
\begin{equation} 
\tilde r(\xi)=-\tilde A\tan (\tilde A\ln\xi-\tilde B);\quad
2\theta(\xi)=\tilde A/\cos (\tilde A\ln\xi-\tilde B)\;,
                                                    \label{e70}     
\end{equation} 
where $\tilde A^2\equiv(2\theta)^2-\tilde r^2$ and $\tan\tilde B\equiv\tilde
r/\tilde A$. The Kondo temperature may be estimated by equating the argument
of  the $\tan$  to $\pi/2$. This way one obtains
\begin{equation} 
T_K=\tilde\Omega \exp\left\{-\left(
{\pi\over 2}+\tilde B\right)/\tilde A\right\}\;.
                                                     \label{e71}   
\end{equation}
The Kondo temperature is maximal for an isolated dot, $\tilde r=0$. In this
case Eq. (\ref{e71}) reduces to \cite{Zwerger93}
$T_K=V^{[0]}\exp\{-\pi^2\sqrt{r_t}/2\}\;$. 
Thus even in the best case the Kondo temperature is exponentially small in the
root of
the tunneling resistance. In order to assure the validity of this theory the
inequality $T_K>\Delta$ should be satisfied, otherwise the assumption about a
continuous spectrum ("metallic dot")
 is not valid and one has to deal with the discrete
spectrum
limit. For a finite value of the environmental resistance, $\tilde r$, the
Kondo temperature rapidly
approaches
zero once the parameters of the system come closer to the separatrix line, Eq.
(\ref{e67}).
This makes the experimental observation of the reentrance behavior quite
difficult 
in practice. 
In contrast, the ``cold'' phase behavior described above seems to be
relatively easy to observe by measuring the temperature dependence of the
heights of the Coulomb blockade peaks.

\section{Summary and Discussion}
\label{s4}

We have addressed here the physics of a (quasi)zero--dimensional quantum
dot
weakly coupled to an ideal lead and capacitively (and not necessarily weakly)
coupled to an external dissipative environment.  The fact that charges on the
dot
interact with the environment which has its own dynamics is translated into an
effective time dependent potential felt on the dot.  In other words, when a
charge
is added to the dot it polarizes the environment. This process is not
instantaneous --- it is characterized by an effective RC time. The delayed
response of
the environment leads to a non--trivial effective time dependent interaction
term.  The (weak) coupling of the lead to the dot means that charges can
tunnel
in and out of the dot, resulting in quantum fluctuations of the charge on the
dot.  We have considered a single channel coupling. For this to
be
the case we require that the linear dimension of the opening connecting the
dot to
the lead does not exceed the Fermi wave length. The methods we employed (mapping 
onto the Coulomb gas and perturbative RG) are restricted to the case of 
relatively weak 
tunneling coupling, that is barrier resistance  larger than $2\pi \hbar/e^2$.
Although we did not restrict ourselves to any finite order in the tunneling 
coupling, a consistent treatment of the strong coupling fixed point is 
outside the scope of this article. The nature of an 
appropriate fixed point was investigated in  Ref. \cite{Matveev95}. 
Our main message here is rather the existence of a separatrix  in 
the parameter (resistance) space of the problem, which separates the basins of 
attraction of
the weak and the strong coupled fixed points.

\begin{figure}
\vglue 0cm
\hspace{0.01\hsize}
\epsfxsize=0.95\hsize
\epsffile{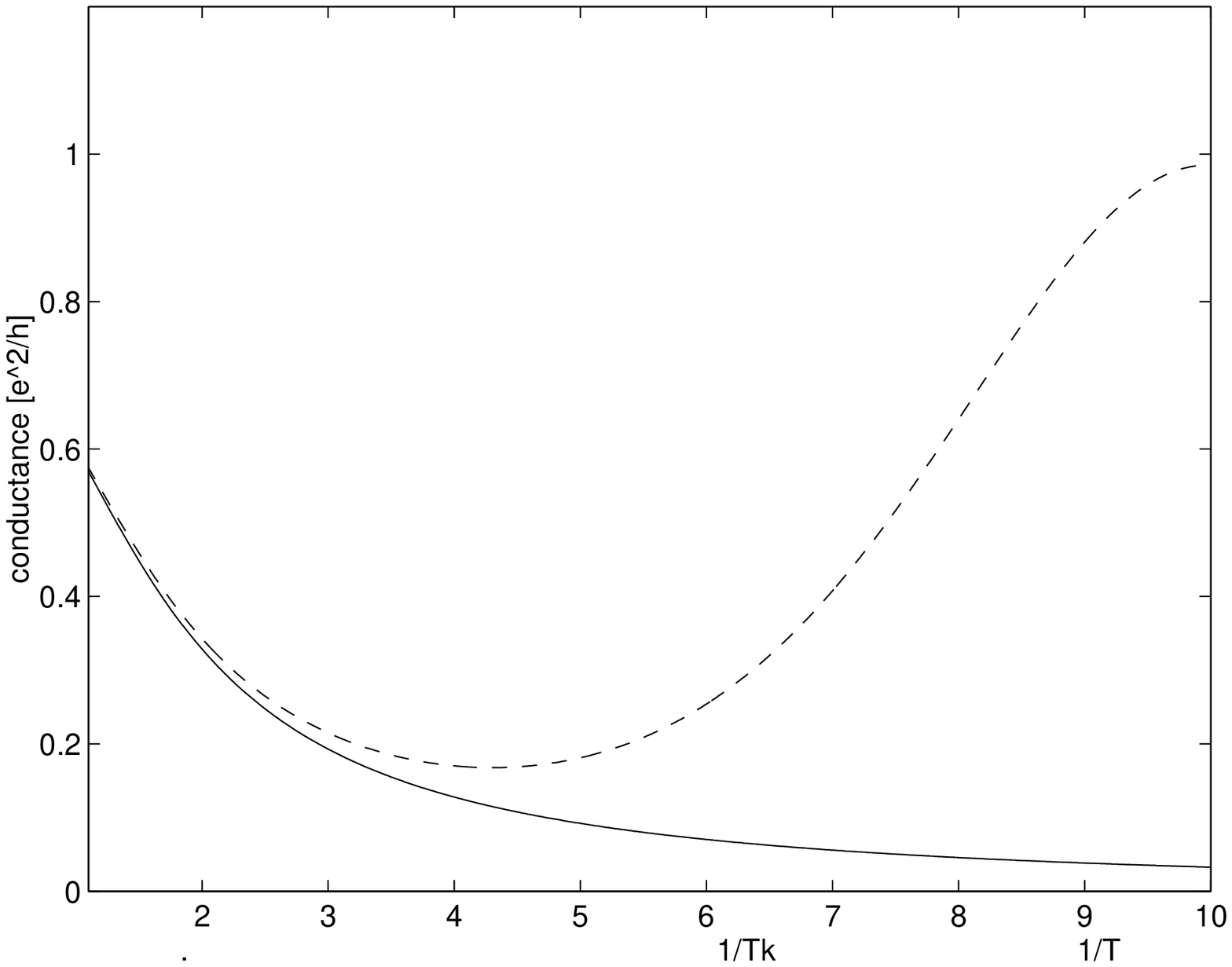}
\refstepcounter{figure} \label{fig4}
{\small FIG.\ \ref{fig4}. Schematic behavior of the conductance (in units 
$e^2/h$) as function of inverse temperature (in units $1/\Omega$) for 
the overdamped (full line) and underdamped (dashed line) scenario; 
$\tilde r=.75$.}
\end{figure}%

We have proceeded by mapping our quantum model onto a classical
one-dimensional neutral Coulomb gas.
The nature of this classical gas  depends crucially
on
whether the Coulomb term is fully or partially screened. In either case we
obtain
the corresponding renormalization group flows. In the former case 
(corresponding to the single tunnel barrier connected to a dissipative 
circuit \cite{Devoret90,Glazman90} ) the flow is towards the ``cold'' phase 
(weak coupling fixed line; Fig. \ref{fig3}a).  
The results of the perturbative treatment  \cite{Devoret90,Glazman90} are 
naturally reproduced in this case. That is, at low temperatures the 
tunneling coupling is suppressed as a certain (positive)
anomalous power of temperature 
(cf. Eq.~(\ref{e60})).  
The case of the dot with a finite residual 
capacitance (partial screening) is more intricate. 
In particular, we found, in agreement with Matveev \cite{Matveev91} and others 
\cite{Zwerger93} that for sufficiently small dissipation of the environment the problem 
is equivalent to a anti-ferromagnetic Kondo. The RG flow is towards the strong 
coupling, characterizing by the ideal transparency of the contact. 
If, however, the resistance of the environment exceeds a certain critical value 
(cf. Eq.~(\ref{e67})) determined by the bare tunneling resistance, the situation 
changes dramatically. The system turns to be overdamped and its low 
temperature behavior is determined by a weak coupling fixed line, similar to 
those of the fully screened situation (with a slightly modified exponent).

The physics of this damping of charge fluctuation by the environment is simple.
Consider a dot which is tuned by the gate electrode
into a resonance, i.e., the cost (in free energy)
of removing one electron from the leads and adding it to the dot is nearly
zero. (We presently assume no voltage difference between the two leads connected
to the dot). The
environmental resistance produces noisy voltage on the gate electrode. Thus, 
at any
given moment, the dot is slightly driven out of resonance. 
Not surprisingly, charge fluctuations are suppressed by this (time--dependent) 
deviation from the exact resonance condition. What is more surprising is an 
existence of a sharp transition at zero temperature
(in fact at $T\leq T_K$ ) between  two 
regimes as a function of the environmental resistance. At a finite temperature 
one has a smooth crossover between these two limits. The
schematic temperature behavior of the tunneling conductance for the two cases 
distinguished by the value of an environmental dissipation is depicted on
Fig. \ref{fig4}. To be within a scenario of an alternating Coulomb gas the 
initial temperature should be less than both $e^2/2(C+C_G)$ and $(RC)^{-1}$.

\par 
Throughout the above analysis it has been assumed that the relation
between the
resistance and the fugacity, Eq. (\ref{e51}), 
remains valid under the renormalization
group  transformation. 
It is possible, though, to establish a direct relation between
the Coulomb gas 
description and the conductance. We first note that the current operator,
${\displaystyle \hat J}$, is given by the time derivative of the number 
operator
${\displaystyle\hat N}=\sum\limits_\alpha a^+_\alpha a_\alpha$ 
\begin{equation} 
\hat J={i\over \hbar}[{\cal H},\hat N]=\Im (W\sum_k d^+_k\sum_\alpha
a_\alpha)\equiv \Im\hat W\;.
                                                   \label{e75}
\end{equation}
The linear conductance may be expressed, employing the Kubo formula, through 
current--current correlation function. Evidently, the latter is
expressible through the two following correlation functions 
\begin{mathletters}
\begin{eqnarray}
&&F(\tau_1-\tau_2)\equiv
\langle{\cal T}\hat W (\tau_1)\hat W^+(\tau_2)\rangle\; , 
                                                    \label{e77}   \\
&&\tilde F(\tau_1-\tau_2)\equiv
\langle{\cal T}\hat W (\tau_1)\hat W(\tau_2)\rangle\; , 
                                                    \label{e78}
\end{eqnarray}
\end{mathletters}
where ${\cal T}$ denotes an (imaginary) time ordering. Both $F$ and $\tilde F$
may be naturally interpreted in a Coulomb gas language. Following the same
procedure
depicted in Appendix B we arrive at the following expression for e.g. 
$F(\tau_1-\tau_2)$ 
\bleq
\begin{equation}
F(\tau_1-\tau_2)={1\over Z}\sum\limits^\infty_{M=0} 
\frac{\theta^{2M+2}}{M!M!}
\int\limits^\beta_0\!\ldots \int\limits^\beta_0\!
d(\Omega\tau_3)\cdots d(\Omega\tau_{2M+2})
\exp\left\{
\sum\limits^{2M+2}_{i<j}(-1)^{i+j}
\left[ 2\ln(\Omega|\tau_i-\tau_j|)+S(\tau_i-\tau_j)\right]\right\}\;.
                                                     \label{e79}
\end{equation}
\eleq
This is the partition function of a one--dimensional classical
neutral gas, the only difference is
that now
two charges of opposite charge are frozen at the points $\tau_1$ and 
$\tau_2$ respectively, all others (whose number is not fixed) 
are free to move. We may
write
$F(\tau_1-\tau_2)\propto\exp\{-U(|\tau_1-\tau_2|)\}$, where $U$ is
the effective screened interaction  between the two frozen opposite charges.
In an analogous way $\tilde F$ may be expressed through the screened 
interaction of two positive (negative) charges. 
It is important to  stress that a single particle propagator (more
generally: a
correlator of an odd number of particles) {\it does not} have a gas
equivalent.
The reason is that different configurations of ``charge'' correspond to terms
of
different signs (an exchange of two charges gives rise to a sign change). Thus
not all terms can be assigned a positive weight.

There are various ways to generalize and extend the model
discussed
here. First, we may include interactions in the lead and even between 
electrons in the lead
and electrons  on the dot. In an obvious notation we denote the two types of
interactions 
by $V_{ll}$ and $V_{0l}$ respectively, while the interaction associated with
the dot
is $V_{00}$. We assume that these (effective) interactions may be time 
dependent, but we do not include
space dependence (only capacitive component). 
The only way these interactions  modify the results is through the definition 
of the effective action, Eq. (\ref{e9}), where one substitutes 
\begin{equation} 
V(\omega)\rightarrow V_{00} +V_{ll}-2V_{0l}\;. 
                                                          \label{e80}
\end{equation}
To see why this is the case one should proceed exactly along the lines of 
Appendices A and B, introducing a doublet of Hubbard--Stratonovich fields, 
$\phi_0, \phi_l$.

\par 
Inclusion of the spin degree of freedom is trivial (redefinition of the
DOS) as long as the tunneling does {\it not} conserve spin. If spin is
conserved in the process of tunneling this is translated into a gas with
charges
(positive and negative) of two ``colors''. Charges of the same color interact
among
themselves via the potential $2\ln\Omega|\tau|+S(\tau)$, as in Eq. (\ref{e50}).
Charges of different colors interact through the potential $S(\tau)$. This
implies
that in the absence of real interaction in the dot ($S=0$), the gas 
factorizes into two mutually non--interacting gases. Using an analogy with
the multichannel Kondo problem one may claim that the separatrix line 
still exists in the system's phase diagram. Its expression through the bare 
parameters of the dot requires, however, some more calculations.

\section{Acknowledgments}

We acknowledge useful discussion with A. Finkelstein,  Y. Oreg and W. Zwerger.
This research was supported by the U.S.-Israel Binational Science Foundation,
the
German-Israel Foundation (GIF) and the Israel Academy of Sciences. Research of 
A.K. was supported by the Rothschild Fellowship.  

\bleq 
\appendix

\section{Single particle Green function} 
\label{a1}

Let us first calculate the 
imaginary time single particle Green function 
of an isolated dot described by a Hamiltonian Eq. (\ref{e1}). It 
may be written as \cite{Negele} 
\begin{equation}
{\cal G}_{\alpha}(\tau_i-\tau_f,\mu)=\frac{1}{Z(\mu)}
\int{\cal D}[\bar a_{\alpha}(\tau)a_{\alpha}(\tau)]
e^{-S[\bar a_{\alpha},a_{\alpha}]  }
\bar a_{\alpha}(\tau_i)a_{\alpha}(\tau_f), 
                                                         \label{4G}
\end{equation} 
with the fermionic action  given by 
\begin{equation}
S[\bar a_{\alpha},a_{\alpha}]= 
\int\limits_0^{\beta}d\tau \left[ \sum_{\alpha}
\bar a_{\alpha}(\tau) 
(\partial_{\tau}+\epsilon_\alpha-\mu) a_{\alpha}(\tau) 
+\frac{V^{[0]}}{2} 
\left[ 
\sum_{\alpha}\bar a_{\alpha}(\tau)a_{\alpha}(\tau) -{\cal N}_0\right]^2 
\right ]; 
                                                         \label{4action}
\end{equation}
here $Z(\mu)$ is the partition function and $\mu$ is the chemical potential. 
Splitting the interaction term in the  action  by means of the 
Hubbard--Stratonovich transformation with 
the  auxiliary Bose field, $\phi(\tau)$, one obtains  
\begin{eqnarray}
{\cal G}_{\alpha}(\tau_i-\tau_f,\mu)=\frac{1}{Z(\mu)}
&&\int{\cal D}[\phi(\tau)]
e^{-\int\limits_0^{\beta}d\tau  \left[ 
\phi(\tau)[2V^{[0]}]^{-1}\phi(\tau) -i {\cal N}_0\phi(\tau) \right] }  
\nonumber \\
&&\int{\cal D}[\bar a_{\alpha}a_{\alpha}]
e^{-\int\limits_0^{\beta}d\tau 
\sum_{\alpha}\bar a_{\alpha} 
(\partial_{\tau}+\epsilon_{\alpha}-\mu+i\phi(\tau)) a_{\alpha} }
\bar a_{\alpha}(\tau_i)a_{\alpha}(\tau_f)     \nonumber \\
=\frac{1}{Z(\mu)}\int{\cal D}[\phi(\tau)]&&
\! e^{-\int\limits_0^{\beta}d\tau  \left[ 
\phi(\tau)[2V^{[0]}]^{-1}\phi(\tau) -i {\cal N}_0\phi(\tau) \right] }
Z^{[\phi]}(\mu){\cal G}^{[\phi]}_{\alpha}(\tau_i,\tau_f,\mu)
                                                          \label{4HS}
\end{eqnarray} 
with the same transformations in $Z(\mu)$. Here $Z^{[\phi]}(\mu)$ and 
${\cal G}^{[\phi]}_{\alpha}(\tau_i,\tau_f,\mu)$ are  respectively 
the partition  
and  Green functions of non--interacting electrons in the time 
dependent (but spatially uniform) potential, 
$i\phi(\tau)$. These quantities may be easily calculated, taking advantage 
of a gauge transformation. The result is 
\begin{equation}
Z^{[\phi]}(\mu)=Z^{[0]}(\mu-i\phi_0),
                                                          \label{4ZG1}
\end{equation}
\begin{equation}
{\cal G}^{[\phi]}_{\alpha}(\tau_i,\tau_f,\mu)=
{\cal G}^{[0]}_{\alpha}(\tau_i-\tau_f,\mu-i\phi_0)
e^{i\int_{\tau_i}^{\tau_f}d\tau [\phi(\tau)-\phi_0]}, 
                                                          \label{4ZG}
\end{equation} 
where $Z^{[0]}(\mu)\equiv\exp\{-\beta\Omega^{[0]}(\mu)\}$ and  
${\cal G}^{[0]}_{\alpha}(\epsilon_n,\mu)=
(i\epsilon_n-\epsilon_{\alpha}+\mu)^{-1}$ are the partition and Green  
functions  of non--interacting electron gas; we have introduced Matsubara 
representation for the boson field, $\phi(\tau)$: 
$$\phi_m\equiv
\beta^{-1}\int_0^{\beta}d\tau\phi(\tau)\exp\{i\omega_m\tau\},$$ 
$\omega_m=2\pi m T$. 
For latter convenience we rewrite 
the exponent in Eq.~(\ref{4ZG}) in the following form 
\begin{equation}
\exp\left\{i\int\limits _{\tau_i}^{\tau_f}d\tau [\phi(\tau)-\phi_0]\right\} =
\exp\left\{
\beta\sum\limits_{m\neq 0}\frac{\phi_{-m} J_{m}^{\tau_i|\tau_f}}{\omega_m} 
\right\}, 
                                                          \label{4exp}
\end{equation} 
where $J_m^{\tau_i|\tau_f}$ is the Matsubara transform of the following 
function 
\begin{equation}
J_{\tau}^{\tau_i|\tau_f}=\delta(\tau-\tau_i)-\delta(\tau-\tau_f).  
                                                          \label{4J}
\end{equation} 
Transforming  
the functional integral over $\phi(\tau)$ to integrals 
over the Matsubara components, $\phi_m$, we obtain 
\begin{eqnarray}
{\cal G}_{\alpha}(\tau_i-\tau_f)=\frac{1}{Z(\mu)}
&&\int d\phi_0
e^{-\beta[\phi_0[2V^{[0]}]^{-1}\phi_0 -
i\phi_0 {\cal N}_0+\Omega^{[0]}(\mu-i\phi_0)]}
{\cal G}^{[0]}_{\alpha}(\tau_i-\tau_f,\mu-i\phi_0)    \nonumber  \\
&&\int\prod_{m\neq 0} d\phi_m
\exp \left\{ \beta  \sum_{m\neq 0} \left[ -\frac{\phi_{-m}\phi_{m}}{2V^{[0]}}
+\frac{\phi_{-m} J_{m}^{\tau_i|\tau_f}}{\omega_m} \right] \right\}. 
                                                          \label{4G1}
\end{eqnarray} 
For   large enough systems ($\Delta\ll T $)  the integral 
over  the static component, $\phi_0$, may be calculated in a saddle point 
approximation  leading to 
${\cal G}^{[0]}_{\alpha}(\tau_i-\tau_f,\overline\mu)$, where 
the stationary point, $\overline\mu$, is the real  solution of the equation 
$(\mu-\overline\mu)/V^{[0]}+N_0+\partial \Omega^{[0]}(\overline\mu)
/\partial\mu=0$. 
The remaining  integrals (over $\phi_m$ for $m\neq 0$) are purely Gaussian. 
As a result one obtains 
\begin{equation}
{\cal G}_{\alpha}(\tau_i-\tau_f,\mu)=
{\cal G}^{[0]}_{\alpha}(\tau_i-\tau_f,\overline\mu)
e^{-S(\tau_i-\tau_f)}, 
                                                          \label{4GT}
\end{equation} 
where 
\begin{equation}
S(\tau)=T\sum_{m\neq 0}\frac{V^{[0]}}{\omega_m^2}
(1-e^{i\omega_m\tau}). 
                                                          \label{4S}
\end{equation} 

We shall include now the environment, Eq~(\ref{e5}), in the calculations of 
the Green function. Assuming that the noise is Gaussian 
with zero mean value, one first perform the averaging over noise realizations 
as   
\begin{equation} 
\langle \ldots \rangle_{noise} = 
\int{\cal D}[\eta(\tau)]
e^{-\frac{1}{2}\int\!\int\limits_0^\beta d\tau d\tau^{\prime} 
\eta(\tau)K^{-1}(\tau-\tau^{\prime})\eta(\tau^{\prime})  }
\ldots\,\, , 
                                                         \label{4aver}
\end{equation} 
where 
$K(\tau-\tau^{\prime})=\langle \eta(\tau) \eta(\tau^{\prime}) \rangle$.  
An important observation is  that the partition 
function of the dot, $Z(\mu)$, is not affected by the noise term \cite{foot2} 
(cf. Eq.~(\ref{4ZG1})).  Averaging Eq.~(\ref{4G}) over 
the Gaussian noise, Eq.~(\ref{4aver}), leads to an effective fermionic action 
with  interaction which is  non--local in time, 
\begin{equation}
S_{int}[\bar a_{\alpha},a_{\alpha}]= \frac{1}{2}
\int \!\!\! \int\limits_0^{\beta}d\tau d\tau^{\prime}  
\sum_{\alpha}\bar a_{\alpha}a_{\alpha} 
\left( V^{[0]}\delta(\tau -\tau^{\prime}) - K(\tau-\tau^{\prime}) \right)  
\sum_{\alpha}\bar a_{\alpha}a_{\alpha}.
                                                         \label{4nlact}
\end{equation}
As a result, one obtains an effective renormalization (screening) 
of the zero--mode interaction potential 
\begin{equation} 
V^{[0]} \rightarrow V(\omega_m)=V^{[0]}- K(\omega_m).
                                                          \label{4Veff}
\end{equation}                        
Further calculations follow  exactly the same steps as outlined above. 
The final result is given by Eqs.~(\ref{e7}) and (\ref{e9}).

\section{Mapping onto a Coulomb gas}
\label{a2}

Our aim here is to derive an expression for the partition
function corresponding to the full Hamiltonian, Eq. (\ref{e6}). 
Performing averaging over the noise and the Hubbard--Stratonovich 
transformation as described in Appendix A one obtains 
\begin{equation} 
Z=\int\!\!{\cal D}[\phi]
\exp\left\{ 
-\sum\limits_{m\neq 0}\frac{\phi_{-m}\phi_m}{2 V(\omega_m)} \right\}
Z[\phi]
                                                             \label{B28}
\end{equation}
where 
\begin{equation}
Z[\phi]=\int\!{\cal D}[\bar a a \bar d d]\exp \left\{
- \int\limits^\beta_0 \! d\tau{\cal L} \right\} \; ,
                                                            \label{B29} 
\end{equation}
\begin{equation} 
{\cal L}=
\sum_\alpha\bar a_\alpha (\partial_\tau+\epsilon_\alpha+i\phi(\tau))
a_\alpha 
+\sum_k\bar
d_k(\partial_\tau +\epsilon_k) d_k+W\sum_{k\alpha}\bar d_k a_\alpha
+W^*\sum_{k\alpha} \bar a_\alpha d_k\;.
                                                            \label{B30}  
\end{equation}
One may integrate out now all fermions of the dot and lead except the two 
living at the point of the tunneling contact by introducing the following 
resolutions of unity  
\begin{equation} 
1=\int\!{\cal D}[s]\delta (s-\sum\limits_\alpha a_\alpha)=\int\!{\cal
D}[s,\bar\sigma] 
\exp\left\{i\int\limits^\beta_0\!d\tau\bar\sigma(s-\sum\limits_\alpha
a_\alpha)\right\}\;,
                                                            \label{B31}  
\end{equation}
and similarly
\begin{equation} 
1=\int\!{\cal D}[r,\bar\rho]
\exp\left\{i\int\limits^\beta_0\!d\tau\bar\rho(r-\sum_k d_k)\right\}\;,
                                                         \label{B32}   
\end{equation}
where $s,r,\bar\sigma,\bar\rho$ are Grassman variables. We also introduce the
two
corresponding identities for the conjugate variables. 
Performing  the
Grassman integration over the original variables $\{a,\bar a,d, \bar d\}$ and 
then over the auxiliary $\{\sigma,\bar\sigma,\rho,\bar\rho\}$ one obtains 
\begin{equation} 
Z[\phi]=Z^{[0]}Z^{[l]}\left \langle \exp\left\{
-\int\limits^\beta_0\!d\tau(W\bar rs +W^*\bar sr)\right\}\right\rangle\;,
                                                         \label{B39}       
\end{equation}
where the angular brackets denote integration over fields $s$ and $r$ with
the following $\phi$-dependent measure:
\begin{equation} 
\det\,{\cal G}^{[\phi]}\det\,{\cal G}^{[l]}\int\!\!{\cal D}[\bar s s \bar r r]
\,  \exp\left\{
-\int\!\!\int\limits^\beta_0 d\tau d\tau'
\big[\bar s({\cal G}^{[\phi]})^{-1}
s+\bar r({\cal G}^{[l]})^{-1}r\big]\right\}\, .
                                                        \label{B40}
\end{equation}
Here we have introduced notations for the traces of the Green functions of
the dot and the lead respectively
\begin{eqnarray}
&&{\cal G}^{[\phi]}(\tau,\tau')\equiv\sum\limits_\alpha\, 
{\cal G}^{[\phi]}_\alpha (\tau, \tau')=
\nu^{[0]} \frac{ \exp\left\{ \beta \sum\limits _{m\ne 0}
{\phi_{-m} J^{\tau|\tau'}_{m}\over \omega_m}\right\} } { \tau-\tau'} \,,
                                               \label{B41a}\\
&&{\cal G}^{[l]}(\tau-\tau')\equiv\sum_k {\cal G}^{[l]}_k(\tau-\tau')
=\nu^{[\ell]}{1\over \tau-\tau'}\;.
                                               \label{B41b}
\end{eqnarray} 
We have assumed that both the dot and the lead have continuous spectra,
characterized by constant density of states ($\nu^{[0]}$ and $\nu^{[l]}$
respectively), and employed Eqs. (\ref{4ZG})--(\ref{4J}). 
At finite temperatures one has to substitute 
\begin{equation}
{1\over\tau-\tau'}\rightarrow{\pi/\beta\over\sin\pi(\tau-\tau')/\beta}\,.
                                                     \label{B43}
\end{equation}
The next step is to expand the exponent in Eq. (\ref{B39}) to infinite order 
in $W$, yielding 
\begin{equation} 
{Z[\phi]\over Z^{[0]}Z^{[l]}}
=\sum\limits^\infty_{M=0}{|W|^{2M}\over (2M)!}{(2M)!\over M!M!}
\int\limits^\beta_0\!\ldots\int\limits^\beta_0\!d\tau_1\ldots
d\tau_{2M}\langle s_1\bar s_2\ldots s_{_{2M-1}} \bar s_{_{2M}}
\rangle\langle\bar r_1
r_2\ldots\bar r_{_{2M-1}}r_{_{2M}}\rangle\; ,
                                                    \label{B44}
\end{equation}
where $s_i\equiv s(\tau_i), i=1,2\ldots 2M$. The combinational factor
$(2M)!/(M!M!)$ represents the number of possibilities to rearrange time
sequences in the angular brackets.  Due to
the Gaussian nature of the measure, Eq~(\ref{B40}), one may apply 
Wick's theorem, resulting
in the $M\times M$ Slatter determinant construction
\begin{equation}
\langle\bar r_1r_2\ldots\bar r_{2M-1}r_{2M}\rangle=\det\big|{\cal G}^{[\ell]}
(\tau_{2n-1}-\tau_{2m})\big|\; ;
                                                    \label{B45}
\end{equation}
$n,m=1,2\ldots M$. We employ now the explicit form of the Green function,
Eq. (\ref{B41b}), along with the known properties of a Cauchy determinant 
\cite{Yuval70} to write
\begin{equation}
\langle\bar r_1 r_2\ldots\bar r_{2M-1} r_{2M}\rangle
=-\big(\nu^{[l]}\big)^M\, {\prod\limits_{n<m}(\tau_{2n}-\tau_{2m})
\prod\limits_{n<m}(\tau_{2n-1}
-\tau_{2m-1})\over\prod\limits_{n,m}(\tau_{2n}-\tau_{2m-1})}\; .
                                                   \label{B46}   
\end{equation}
In the same manner 
\begin{eqnarray}
\langle s_1\bar s_2\ldots s_{2M-1}\bar s_{2M}\rangle 
&= -\big(\nu^{[0]}\big)^M\, 
{\prod\limits_{n<m}(\tau_{2n}-\tau_{2m})\prod\limits_{n<m}(\tau_{2n-1}-\tau_{2m-1})
\over \prod\limits_{n,m}(\tau_{2n}-\tau_{2m-1})}\nonumber\\
&\exp\left\{\beta\sum_{m\ne 0}{\phi_{-m}
J_{m}^{\tau_2\ldots\tau_{2M}|\tau_1\ldots\tau_{2M-1}}\over\omega_m}
\right\}\; .
                                                   \label{B47} 
\end{eqnarray}
where $J_{m}$ is the Fourier transform of the imaginary time function 
\begin{equation} 
J^{\tau_2\ldots\tau_{2M}|\tau_1\ldots\tau_{2M-1}}_\tau 
=\sum\limits^{2M}_{j=1}
(-1)^j\delta(\tau-\tau_j)\;.
                                                  \label{B48}
\end{equation}
The exponent in Eq. (\ref{B47}) is the only part of $Z[\phi]$ which depends
on the Hubbard--Stratonovich field, $\phi(\tau)$. Thus integration over the
field $\phi$, Eq. (\ref{B28}), may be easily performed for each term in the 
sum (Eq. (\ref{B44})) separately. Indeed
\begin{equation}
\int\!\!{\cal D}[\phi_m]
\exp\left\{-\beta\sum\limits_{m\ne 0} \left[ 
{\phi_{-m}\phi_{m}\over 2V(\omega_m)} - 
{\phi_{-m} J_{m}\over \omega_m} \right] \right\} =
\exp\left\{-\beta\sum\limits_{m\ne 0} 
J_{-m}{V(\omega_m)\over 2\omega_m^2}J_{m}\right\}\; .
                                              \label{B49}
\end{equation}
One may now return to time representation and employ the definition of
$J_\tau$, Eq. (\ref{B48}), as well as Eq. (\ref{e9}) to write the exponent 
on the r.h.s. of Eq. (\ref{B49}) as 
$\sum\limits^{2M}_{i<j}(-1)^{i+j} S(\tau_i-\tau_j)$. 
Note
also that the product of the two Cauchy determinants (Eqs. (\ref{B46}) and 
(\ref{B47})),
being positive definite, may be written as an exponent of
$\sum\limits^{2M}_{i<j}(-1)^{i+j}\, 2\ln|\tau_i-\tau_j|$. 
Finally one obtains Eq.~(\ref{e50})
for the partition function of the dot connected to the lead. 
\eleq

\bleq
\begin{center}{\bf Table 1}\end{center}
\begin{table}[t]
\begin{center}
\begin{tabular}{lcc}\hline
&&\\
&Quantum Dot&Coulomb Gas\\ 
 &&\\ \hline
&&\\ 
(temperature)$^{-1}$ & $\beta$ & $2+2\tilde r$\\
&&\\
linear size & L& $\beta$\\
&&\\
dimension & ``0"+1 & 1\\
&&\\
potential& $V(\omega)$ & $\ln (\Omega|\tau|)$ \\
&&\\
fugacity & $e^{\mu\beta}$ & $\theta$\\ 
&& \\ \hline
\end{tabular}\end{center}\end{table}

\eleq

\ecols

\begin{thebibliography}{34}
\bibitem{foot1}
It is interesting to note that $V(\omega_m)$ may also be obtained
through a standard evaluation of the polarization operator. 
For a diffusive system
where coupling to the environment is described by a constant leakage rate
$\gamma_{in}$, such a dynamically screened (and $q$ dependent) interaction
can be written as
$$V(q,\omega_m)=\left[{1\over V^{(0)}(q)}+{dn\over d\mu}\;{Dq^2
+\gamma_{in}\over Dq^2 +|\omega_m|+\gamma_{in}}\right]^{-1}\;. $$  
Here $dn\over d\mu$ is the thermodynamic density of states 
(compressibility), $V^{(0)}(q)$ is the unscreened interaction. At 
$q$=0 and $\omega\gg\gamma_{in}$ this equation reduces to Eq. (\ref{e16}) 
with $C_G=\infty\, , V^{(0)}(q=0)=e^2/ C\, ,$ and $R=\Delta/(2\gamma_{in})\,$
($\Delta$ being the mean level spacing).
\end{thebibliography}
\end{document}